# Strong collisionless damping of the low-velocity branch of electromagnetic wave in plasmas with Maxwellian-like electron velocity distribution function[1]

V. N. Soshnikov [2]

Plasma Physics Dept.,
All-Russian Institute of Scientific and Technical Information
of the Russian Academy of Sciences
(VINITI, Usievitcha 20, 125315 Moscow, Russia)

**Abstract**

*After approximate replacing of Maxwellian distribution exponent with the rational polynomial fraction we have obtained precise analytical expression for and calculated the principal value of logarithmically divergent integral in the electron wave dispersion equation. At the same time our calculations have shown the presence of strong collisionless damping of the electromagnetic low-velocity (electron) wave in plasmas with Maxwellian-like electron velocity distribution function at some small, of the order of several per cents, differences from Maxwellian distribution $\sim \exp(-\xi^2)$ in the main region of large densities $\xi^2 \lesssim 4$, where $\xi^2$ is dimensionless energy parameter, however due to the differences in the tail with $\xi^2 > 4$, where electron density itself is negligibly small.*

**Introduction**

We have argued and used in the preceding works [1 – 4] assertion that logarithmically divergent integrals in the dispersion equation of electromagnetic waves in plasma ought to be understood (taken) in the principal value sense, what entails absence of collisionless damping of waves in homogeneous plasma with Maxwellian electron energy distribution function. It is directly verified by the before found non-damping finite solutions for the problem of waves excited by the boundary periodic electrical field $E_0 \exp i\omega t$, and follows from a simple postulate that the appearance itself of indefinitely divergent integrals means the necessity of accounting for some lacking additional reasonable physical conditions which, as it was found, can be satisfied only when such integrals have the principal value sense [1].

The difficulties of calculating the principal sense integrals can be avoided using approximations of integrand exponent, replacing the latter with a fitted rational fraction, so after this the integrals can be taken analytically. However approaching to the exponent with this approximate function can lead to strong collisionless damping of the low velocity electron wave due to, whereas one could think it to be insignificant, differences of small electron densities from Maxwellian ones in the distribution tail. On the contrary, the high velocity branch with near light velocity of propagation of electromagnetic waves is weakly depending on the mean electron velocity in difference from the low-velocity wave, since the damping in the former can appear only as small terms of the order $\overline{v_x^2}/c^2$, where $x$ is direction of wave propagation, $v_x$ is electron velocity, $c$ is velocity of light.

---

Since the real distribution functions differ from Maxwellian ones, we propose in this paper to account for these differences approximating distribution function with a symmetric relative to $\pm\xi$ rational fraction $1/f$ where

$$f = \sum_{n=0}^{N} a_n \xi^{2n}, \qquad a_0 = 1 \qquad (1)$$

with free parameters $a_n$ for the possibly best approach to either real distribution function or Maxwellian one $1/f \simeq \exp(-\xi^2)$ as the limiting case. However the precision of this procedure augments only with increasing a number of free parameters $N$, what leads to growing the polynomial power and corresponding solving difficulties of algebraic dispersion equation with a necessity of analysis and possible neglecting accessory roots. Nevertheless this methodology appears viable and is further demonstrated with the simplest examples with $N = 2$ and $N = 3$ which characterize main features of the low-velocity electromagnetic waves in plasmas. At the same time the increasing of $N$ leads to, as it is illustrated further, to the necessity of finding complex roots of algebraic equations of the order $N \geq 6$, what is possible practically only with using computer technologies.

**Replacing of exponent**

If $x$ is coordinate in the direction of wave propagation and electrical field is directed along $z$, for harmonic transverse electrical field at the boundary of a half-infinite homogeneous electron plasma slab we have standard dispersion equation of electromagnetic waves (see also [3])

$$G \equiv p_2^2 - \frac{p_1^2}{c^2} + \frac{\omega_L^2 p_1}{c^2} \int v_z \frac{\partial f_0}{\partial v_z} \frac{d\vec{v}}{p_1 + v_x p_2} = 0, \qquad (2)$$

where $p_1 \equiv i\omega$; $p_2 \equiv -ik$; $\mathrm{Re}\, k > 0$ are parameters of Laplace transform in asymptotical wave solution for the forward wave

$$E_z(x) \equiv E \propto \exp(i\omega t - ikx). \qquad (3)$$

Here $\omega_L$ is Langmuir electron frequency $\omega_L = 4\pi e^2 n_e / m_e$; $c$ is light velocity in vacuum; $f_0$ is background electron energy distribution function normalized to unity. In the case of Maxwellian distribution function

$$f_0 = \left(\frac{m_e}{2\pi k_B}\right)^{3/2} e^{-\frac{m_e v^2}{2 k_B T}}, \qquad (4)$$

where $k_B$ is Boltzmann constant; $T$ is temperature; $m_e$ is mass of electron; $v$ is electron velocity $|\vec{v}|$; integration in $v_z$, $v_y$ in Eq. 2 leads to expression

$$\frac{\omega_L^2 p_1}{c^2} \int v_z \frac{\partial f_0}{\partial v_z} \frac{d\vec{v}}{p_1 + v_x p_2} = -\frac{\omega_L^2}{c^2} \frac{\beta}{\sqrt{\pi}} \int \frac{e^{-\xi^2}}{\beta - \xi} d\xi, \qquad (5)$$

where

$$\beta \equiv \frac{\omega}{k}\sqrt{\frac{m_e}{2k_B T}}; \quad \text{Re}\,\beta > 0. \qquad (6)$$

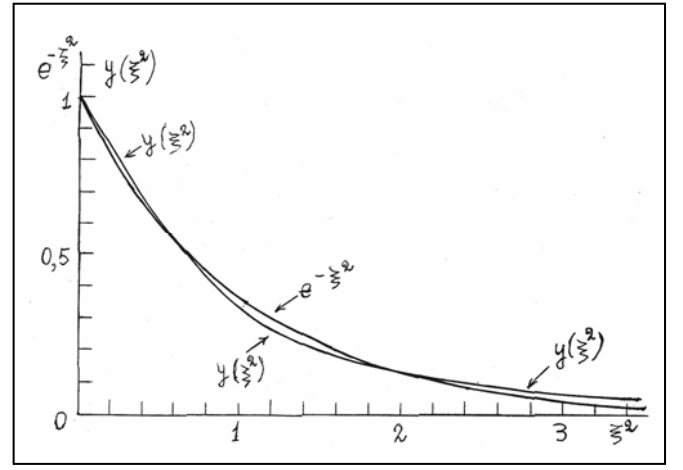

In the case of normalized Maxwellian-like distribution function of the type $f_0 = f_x(v_x) f_y(v_y) f_z(v_z)$ integration in $v_y$, $v_z$ leads to appearance of numerical factors in the integral (5) which are easily accounted for if it is necessary at the following solution of dispersion equation, so further these complications are not considered.

*Рис. 1. Approximation of exponent (non-optimized variant) $exp(-\xi^2) = y(\xi^2) = 1/f(\xi^2) = 1/(1.233\xi^4 + 0.739\xi^2 + 1)$*

Fig. 1 illustrates a non-optimized variant $N = 2$ of approximation of exponent $\exp(\xi^2)$ with rational fraction $1/f = 1/(a\xi^4 + b\xi^2 + 1)$; $a = 1.233$; $b = 0.739$ up to the boundary $\xi^2 \lesssim 3$ of the main region of large electron densities. In this case normalization integrals also approximately agree:

$$N_{\exp}^{-1} = \int e^{-\xi^2} d\xi = \sqrt{\pi} \simeq 1.772 \qquad (7)$$

and

$$N_{polin}^{-1} = \int \frac{d\xi}{a\xi^4 + b\xi^2 + 1} = \frac{\pi}{a^{1/4}} \frac{\sin(\Psi/2)}{\sin \Psi} \simeq 1.83, \qquad (8)$$

where

$$\Psi = \arctan \frac{\sqrt{\frac{1}{a} - \frac{b^2}{4a^2}}}{b/2a} \simeq 70.5°. \qquad (9)$$

We have correspondingly for the mean square dimensionless velocities $\sqrt{\overline{\xi^2}}$

$$\overline{\xi^2_{\exp}} = \frac{1}{\sqrt{\pi}} \int \xi^2 e^{-\xi^2} d\xi = 0.5; \qquad (10)$$

$$\overline{\xi^2_{polin}} = N_{polin} \int \frac{\xi^2 d\xi}{a\xi^4 + b\xi^2 + 1} = N_{polin} \frac{\pi}{a^{3/4}}\left(\cos\frac{\Psi}{2} - \frac{b}{2a^{1/2}} \frac{\sin(\Psi/2)}{\sin \Psi}\right) \simeq 0.90. \qquad (11)$$

Such a considerable discrepancy can be removed only by increasing of power $N$ in the rational fraction which approximates exponent. Nevertheless it appears useful to calculate further contribution of the relatively small part of high energy electrons upon dispersion features of the low-velocity electron wave.

Representing the rational fraction as a sum of more simple fractions one obtains after some elementary but enough cumbersome transformations

$$-\frac{\omega_L^2}{c^2}\frac{\beta}{\sqrt{\pi}}\int\frac{d\xi}{\left(a\xi^4+b\xi^2+1\right)(\beta-\xi)}=-\frac{\omega_L^2}{c^2\sqrt{\pi}}\frac{\beta F(\beta)}{a\beta^4+b\beta^2+1}, \quad (12)$$

where

$$F(\beta)=\int\frac{a\xi^2+a\beta^2+b}{a\xi^4+b\beta^2+1}d\xi+\int\frac{d\xi}{\beta-\xi}, \quad (13)$$

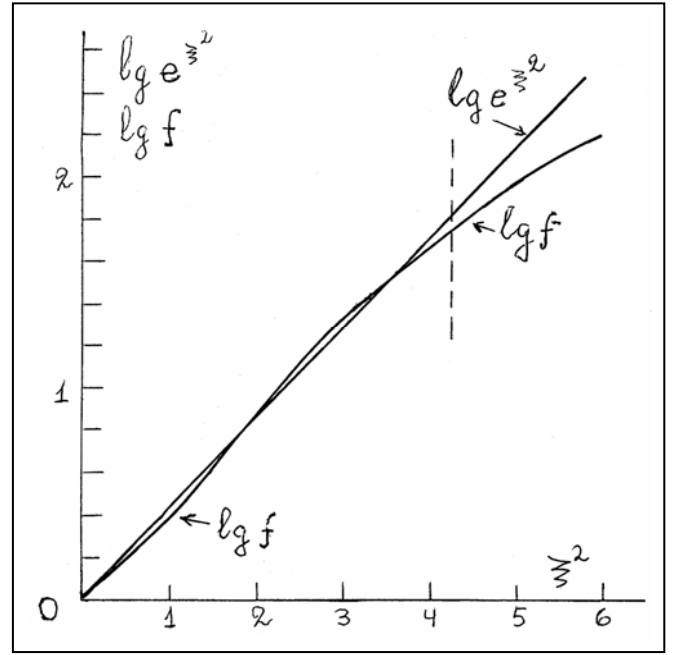

Рис. 2. Approximation of exponent (non-optimized variant)
$exp(\xi^2)=f=0.822\xi^6+0.649\xi^4+1.36\xi^2+1$

and these integrals with symmetrically tending integration limits $\xi\to\pm\infty$ can be taken analytically and calculated precisely with some further reserves about the second indefinite integral term in (13).

A non-optimized variant of the more precise approximation of the exponent at $N=3$ is illustrated with Fig. 2, where relatively good approach has place up to values $\xi^2\lesssim(4-4.5)$.

Calculations of some selected variants $N=5$ show that acceptable approximation with an error less than $(5-10)\%$ can be expected up to $\xi^2\lesssim 10$ (when $\exp(-10)=4.54\times10^{-5}$) already at $N=6$ at condition of using optimized coefficients $a_n$.

**Solutions of dispersion equation for low-velocity wave**

For low-velocity branch one can neglect in Eq.2 terms of the order $\sim\overline{v_x^2}/c^2$ that is term $p_1^2/c^2$ in comparison with $p_2^2=-k^2=\dfrac{-\omega^2}{\beta^2}\dfrac{m_e}{2k_BT}$ (in analogy with that in [3]). Then Eq.2 reduces to equation

$$-\frac{\beta^3}{\sqrt{\pi}}\frac{\omega_L^2}{\omega^2}\frac{2k_BT}{m_ec^2}\frac{F(\beta)}{a\beta^4+b\beta^2+1}=1. \quad (14)$$

Following this one can obtain approximate solution of Eq.14 supposing as it was made in [3] at $p_2=(p_2)_0+\sigma$ with small $\sigma$

$$\beta=\beta_0\left(1+\frac{\tau}{\beta_0}\right), \quad (15)$$

where

$$|\tau|\ll|\beta_0|, \quad (16)$$

and $\beta_0$ satisfies equation

$$a\beta_0^4+b\beta_0^2+1=0. \quad (17)$$

Substituting (15) into (16) one obtains equation for finding $\tau$

$$\tau \approx -\frac{\beta_0^4}{\sqrt{\pi}} \frac{\omega_L^2}{\omega^2} \frac{k_B T}{m_e c^2} \frac{F(\beta_0)}{2a+b}, \qquad (18)$$

where according to (17)

$$\beta_0^2 = -\frac{b}{2a} \pm i\frac{\sqrt{4a-b^2}}{2a}. \qquad (19)$$

Note, that if the before removed term $p_1^2/c^2$ in Eq. 2 would be kept, in Eq.18 it would take the form $p_1^2 \left(\overline{v_x^2}/c^2\right)(\tau/\beta_0)$, what confirms possibility of its neglecting.

Representing $\beta_0$ in canonical form

$$\beta_0 = \rho \exp(i\varphi) \qquad (20)$$

one obtains

$$\rho = a^{-1/4}; \quad 2\varphi = \arctan \frac{\pm\sqrt{4a-b^2}/2a}{-b/2a}. \qquad (21)$$

Negative sign of the square root in Eq. 21 corresponds to angle $2\varphi$ in the third quadrant of the circle, that is angle $\varphi$ is placed in the second quadrant where $\mathrm{Re}\,\beta_0 < 0$. For to have $\mathrm{Re}\,k_0 > 0$, correspondingly $\mathrm{Re}\,\beta_0 > 0$, one ought to select according to (21)

$$\varphi = \frac{\pi}{2} - \frac{\Psi}{2}, \qquad (22)$$

where $\Psi$ is determined by Eq. 9. Then $\mathrm{Im}\,\beta_0 > 0$, what corresponds to damping wave. But in this case the second integral in Eq.13 with complex value $\beta_0$ and $\mathrm{Re}\,\beta_0 > 0$ equals

$$\int \frac{d\xi}{\beta-\xi} = -i\pi, \qquad (23a)$$

since in presentation $\xi \equiv \rho_\xi \exp(i\varphi_\xi)$ we could write the following sequence of elementary relations:

$$\int \frac{d\xi}{\beta_0-\xi} = -Log(\beta_0-\xi)\Big|_{-\xi \to -\xi_1 \to -\infty}^{\xi \to \xi_1 \to \infty} \to -Log\,\frac{-\xi_1}{+\xi_1}\Big|_{-\infty}^{+\infty} = -i\pi \qquad (23b)$$

(as selected from one of values $Log(-1) = \pm i\pi$).

However there is also an alternative probable variant of real value $\beta_0$ and correspondingly real value integral with usually taken principal sense of indefinite integral in (23a) as

$$\int_{-\xi_1\to-\infty}^{\xi_1\to\infty} \frac{d\xi}{\beta-\xi} = 0, \qquad (23c)$$

what changes the value of small addition term $\tau$ in Eq. 18.

In spite of this in the following calculation of $\tau$ we have taken arbitrarily the expression (23a), but the final choice can be justified only with some additional physical considerations. So, in the case of half-infinite plasma slab these ones are the absence of kinematical and fast backward electromagnetic waves, and also possible other physical conditions with the proposed absence of wave damping in the Maxwellian distribution limit of large $N$ with using the integral value (23c) as soon as relations (23a) and (23b) are equivalent to contour integrals along the straight coordinate line $\text{Im}\,\xi = \text{Im}\,\beta$ in the complex value plane $\xi$ with the prescribed by Landau theory passing around the pole $\xi = \beta$, and this sense of indefinite integral contradicts to the spirit and statements of the work [1].

Cumbersome but elementary calculations of the first integral in Eq. 13 lead to expression

$$\int \frac{a\xi^2 + a\beta_0^2 + b}{a\xi^4 + b\xi^2 + 1} d\xi = \frac{\pi}{a^{1/4}}\left(\frac{b}{2} + a\beta_0^2\right)\frac{\sin(\Psi/2)}{\sin\Psi} + \pi a^{1/4}\cos\frac{\Psi}{2} \simeq 2.25\beta_0^2 + 3.39 \simeq 2.71 + 1.91i, \qquad (24)$$

where

$$\beta_0 = \frac{1}{a^{1/4}}\left(\sin\frac{\Psi}{2} + i\cos\frac{\Psi}{2}\right) \simeq 0.547 + 0.775i; \qquad (25)$$

$$\beta_0^2 = \frac{1}{a^{1/2}}(-\cos\Psi + i\sin\Psi) \simeq -0.301 + 0.849i; \qquad (26)$$

$$\beta_0^4 = \frac{1}{a}(\cos 2\Psi - i\sin 2\Psi) \simeq -0.626 - 0.511i, \qquad (27)$$

since angle $(-2\Psi)$ finds oneself in the third quadrant of the circle, and angle $(+2\Psi)$ in the second one. Hence

$$k_0 = \frac{\omega}{\sqrt{\overline{v_{x\exp}^2}}}(0.432 - 0.611i)\left[1 - \frac{\omega_L^2}{\omega^2}\frac{\overline{v_{x\exp}^2}}{c^2}(0.342 - 0.289i)\right], \quad (\overline{v_{x\exp}^2} = \frac{k_B T}{m_e}). \qquad (28)$$

Besides the variant illustrated by Fig. 1 there is considered also a non-optimized variant of Fig. 2 with $N = 3$.

Replacing Maxwellian exponent approximately with a rational fraction analogously to the foregoing, one obtains cubic relatively to $\beta_0^2$ equation for determination $\beta_0$, and then correspondingly $k_0$

$$0.822\beta_0^6 - 0.649\beta_0^4 + 1.366\beta_0^2 + 1 = 0, \qquad (29)$$

which was solved according to trigonometric formulas of [5]. The final solution for forward wave was determined corresponding to condition $\text{Re}\,k_0 > 0$ with suitable selection of the roots, and has the form

$$\beta_0 = 0.913 + 0.723i; \tag{30}$$

$$k_0 = \frac{\omega}{\sqrt{\overline{v^2_{x\exp}}}}(0.476 - 0.466i), \qquad \left(\overline{v^2_{x\exp}} = \frac{k_B T}{m_e}\right). \tag{31}$$

Comparing expressions (28) and (31) with the proposed before in [1 – 4] value $v_{xeff} \simeq \sqrt{\overline{v_x^2}}$ one can conclude that one ought to precise the latter by its factorizing up to $\sim (1.5 - 2)$ times.

Note, that in the case of Maxwellian-like distribution function corrections to $k$ of the high-velocity wave are very small, so they can be calculated with linearization $k \to k_0 + \sigma$, $|\sigma| \ll |k_0|$ in dispersion equation with corresponding considerable simplifications in approximate taking integral of derivatives of the rational fraction $1/f(\xi^2)$.

**Discussion**

Since the rational fraction $1/f(\xi^2)$ approximates exponent $\exp(-\xi^2)$, the roots of polynomials of the type (17) and (29) (corresponding to equation $f(\xi^2) = 0$) for finding $\beta_0^2$ can not all be positive real values, and the paradox of the complex-value roots which lead to damping waves is resolved with the fact that at increasing polynomial power $N$ imaginary parts of the corresponding roots $\beta_0^2$, $\beta_0$ and also $k_0$ must quickly decrease.

The principal value of logarithmically divergent integral in the dispersion equation of plasma electron waves can be calculated analytically at approximate replacing integrand exponent $\exp(-\xi^2)$ with a rational fraction $1/\sum_{n=0}^{N} a_n \xi^{2n}$, $a_0 = 1$, with optimally fitted constants $a_n$. The value $N = 2$ is sufficient for agreement of normalization factors (that is electron densities), however mean square values $\overline{\xi^2}$ diverge up to 2 times. The precision of approach must improve at increasing $N$, but then difficulties augment related with finding roots of high power algebraic denominator of dispersion equation and both analyzing and possibly neglecting accessory roots, since in the limit of Maxwell distribution at large $N$ there exist only one wave branch for longitudinal waves and only two wave branches for transverse waves, therefore less important other also existing branches which correspond to all totality of roots $k_0$ were not considered here. In our case the question is about solution of algebraic $N$-power part of dispersion equation of the type Eqs. (17) and (29).

Demonstrative calculations at $N = 2$ and $N = 3$ for electromagnetic waves show that declination of electron distribution function from Maxwellian one affects very weakly the high-velocity wave branch whereas the low-velocity wave branch suffers strong damping on account for surplus of non-Maxwellian fast electrons at $\xi^2 \gtrsim 4$, besides that real value part of the wave number decreases, what can be regarded as a more precise definition of the before taken values $\operatorname{Re} k$ and $v_{xeff}$ in papers [1 – 4].

In spite of the above-mentioned difficulty of numerical finding all roots $\beta_0^2$ and $\beta_0$ of the dispersion equation and their analysis and selection of all suitable variants, by us obtained results show the possibility of applying this method in the case of Maxwellian-like and other background distribution functions. However the more precision results require $N \geq 6$ what is possible practically only applying computer technologies using programs of optimizing constants $a_n$, numerical calculation of integrals non-comprising $\beta_0$ in their integrand functions and numerical solution of algebraic relatively to $\beta_0^{2n}$ equations with preliminary calculated numerical coefficients.

According to the foregoing, solution of the collisionless dispersion equation can be obtained by replacing in integrand functions $v_x \to v_{xeff} = \bar{v}_x \gamma_{x1}$ where $\gamma_{x1}$ is dimensionless parameter needed to determination, and $\bar{v}_x \equiv \sqrt{2/\pi}\sqrt{k_B T/m_e}$.

Analogous procedure of replacing

$$v_x \to \bar{v}_x \gamma_x, \quad v_y \to \bar{v}_y \gamma_y, \quad v_z \to \bar{v}_z \gamma_z \tag{32}$$

can be made also at calculation of integrals determining contribution of Coulomb collisions in the first iterative approximation.

Using expression for collision term in [3] with differentiations in the integrand and using replacements (32) one obtains after some cumbersome transformations the following renewed results for derived in [3] decrements $\delta$ of collision damping:

$$\delta_1 = -\frac{2\pi e^4 n_i Z^2 L \omega_L^2}{m_e c k_B T \omega^2} \frac{1}{3\sqrt{3\overline{v_x^2}}} \frac{1}{\sqrt{1-\omega_L^2/\omega^2}} F_1 \tag{33}$$

- at fast electromagnetic waves;

$$\delta_2 = -\left(\frac{\pi e^4 n_i Z^2 L \omega^2}{3\sqrt{3}\left(\overline{v_x^2}\right)^2 m_e k_B T}\right)^{1/3} F_2^{1/3} \tag{34}$$

- at low-velocity transverse electron waves, and

$$\delta_3 = -\frac{2\pi e^4 n_i Z^2 L \omega^4}{3\sqrt{3}\left(k_B T\right)^2 \omega_L^4 \sqrt{1-\omega_L^2/\omega^2}} F_3 \tag{35}$$

- at longitudinal electrostatic low-velocity waves, where accounting for differences of values ($\gamma_{x1}$, $\gamma_x$, $\gamma_y$, $\gamma_z$) in either $F_1$, $F_2$, $F_3$

$$F_1 = \left(\frac{3}{2}\right)^{3/2} \sqrt{\pi} \frac{\gamma_x^2 + \gamma_y^2}{\left(\gamma_x^2 + \gamma_y^2 + \gamma_z^2\right)^{3/2}}, \tag{36}$$

and $\gamma_x$, $\gamma_y$, $\gamma_z$ - are constants independent of $\beta$;

$$F_2 = \frac{\pi^2\sqrt{3}}{16} \frac{1}{\left(\gamma_x^2 + \gamma_y^2 + \gamma_z^2\right)^{3/2}} \frac{1}{\gamma_{x1}^5 \gamma_x^4} \left(\gamma_{x1}^4 \gamma_y^2 \gamma_z^2 + \gamma_{x1}^4 \gamma_z^4 - 6\gamma_{x1}^4 \gamma_x^2 \gamma_z^2 + \gamma_x^4 \gamma_y^2 \gamma_z^2 + \gamma_x^4 \gamma_z^4 + 3\gamma_x^6 \gamma_z^2 + 6\gamma_{x1}^2 \gamma_x^2 \gamma_y^2 \gamma_z^2 + \right.$$
$$\left. + 6\gamma_{x1}^2 \gamma_x^2 \gamma_z^4 + 3\gamma_{x1}^2 \gamma_x^4 \gamma_z^2 - \gamma_{x1}^6 \gamma_x^2 - \gamma_{x1}^6 \gamma_y^2 - \gamma_{x1}^2 \gamma_x^6 - \gamma_{x1}^4 \gamma_x^4 - \gamma_{x1}^4 \gamma_x^2 \gamma_y^2 + \gamma_x^8\right), \tag{37}$$

where $\gamma_x$, $\gamma_y$, $\gamma_z$ depend only on the value $\gamma_{x1}$;

$$F_3 = \frac{3\sqrt{3}\pi}{4\gamma_{x1}} \frac{\gamma_y^2 + \gamma_z^2}{\left(\gamma_x^2 + \gamma_y^2 + \gamma_z^2\right)^{3/2}} \left(1 + \frac{\gamma_x^4}{\gamma_{x1}^4} - 2\frac{\gamma_x^4}{\gamma_{x1}^4}\frac{\omega_L^2}{\omega^2} + \frac{\gamma_x^4}{\gamma_{x1}^4}\frac{\omega_L^4}{\omega^4} + 6\frac{\gamma_x^2}{\gamma_{x1}^2} - 6\frac{\gamma_x^2}{\gamma_{x1}^2}\frac{\omega_L^2}{\omega^2}\right), \tag{38}$$

where $\gamma_x$, $\gamma_y$, $\gamma_z$ depend only on $\gamma_{x1}$ and the factor $\left(1 - \omega_L^2/\omega^2\right)^{-1/2}$ appearing in $\beta$ after substitution therein the value $k$.

At the end it ought to note that there remains unclear the role of appearing numerous wave branches corresponding to polynomial fraction representing electron distribution function.

**Conclusion**

We have demonstrated possibilities of approximate representation of exponent $\exp(-\xi^2)$ by its replacing with rational polynomial fraction $1/f(\xi^2)$ what allows analytical calculation of the principal values of logarithmically divergent integrals in dispersion equations of plasma waves and also their numerical calculations due to detachment and casting off indefinitely divergent part $\int \frac{dx}{a-x}$.

It is shown that there arises strong collisionless damping of the low-velocity electron wave due to non-Maxwellian tail of fast electrons at $\xi^2 \sim (4-5)$.

Solving dispersion equation for finding parameters of the low-velocity branch is reduced in the first approach to finding corresponding roots $\beta_0$ of algebraic equation $f(\beta_0^2) = 0$ with the following calculation of the wave numbers $k_0 \propto 1/\beta_0$.

Due to the smallness of the following approaches these ones can be obtained in all cases with linearization $k = k_0 + \sigma$, $|\sigma| \ll |k_0|$.

Proposed in the works [1 – 4] value of effective velocity $v_{xeff} \simeq \sqrt{v_x^2}$ at evaluation of integrals ought to be increased up to $\sim (1.5 - 2)$ times.

For practical results the calculation precision ought to be increased with a good polynomial approximation of exponent up to values $\xi^2 \gtrsim 10$ according to power of approximating polynomials in (1) $N \geq 6$. This is attainable only at applying computer technologies.

Use of the method of constant effective values $v_{xeff}$ of integrand variable $v_x$ results in a statement that the before found expressions for collision damping decrements can be corrected by factorization with some universal constants except longitudinal waves for which these coefficients comprise also dependence on relation $\omega_L^2/\omega^2$.

Since collisionless damping is consequence of non-Maxwellian type of electron background velocity distribution function it appears expedient to consider also possible damping in the case of distribution functions at most different from Maxwellian ones, par example, $\delta$-like distributions. Such investigation for one- and two-stream plasma was undertaken in our papers [6], [7] with analysis of existing solutions both damping/growing type and non-damping ones with superposition of inherent plasma electron and boundary field excitation frequencies. It ought to note intriguing absence in the case of $\delta$-type distributions any problems connected with the indefinitely divergent integrals.